# 800°C Air-Ambient Operation of Extreme-Bandgap $Al_xGa_{1-x}N$ (x>0.6) Digital and Random-Alloy Channel HEMTs

Abdullah Al Mamun Mazumder[1], Tariq Jamil[1], Mafruda Rahman[1], Muhammad Ali[1], Ankit Malik[1], Anthony Joyce[1], Patrick Darmawi-Iskandar[2], Chandan Joishi[3], Mansura Sadek[3], James G. Fiorenza[3], Manoj Kumar Chandra Mohan[1], Grigory Simin[1], Tomás Palacios[2], and Asif Khan[1]

[1]Department of Electrical Engineering, University of South Carolina, Columbia, SC 29208, USA

[2] Massachusetts Institute of Technology, Cambridge, MA 02139, USA

[3]Analog Devices Incorporated, Boston, MA 02110, USA

**Authors to contact:** mazumdea@email.sc.edu

**This work demonstrates the high-temperature (room to 800 °C) operation of thin $Al_{0.62}Ga_{0.38}N$ digital alloy (DA) short-period superlattice (SPSL), and random alloy (RA) $Al_{0.64}Ga_{0.36}N$ channel MOSHEMTs on sapphire. The RA devices employ a doped $n^+$-85%-Al barrier and a highly doped $n^{++}$-$Al_{0.86\rightarrow0.20}Ga_{0.16\rightarrow0.80}N$ reverse-graded layer (RGL) for surface ohmic contact formation, whereas the DA devices employ a binary AlN barrier and selectively regrown $n^{++}$-GaN source/drain contacts. At 800 °C, the DA channel devices exhibit better thermal stability: their saturated drain current decreases from 460 to 172 mA/mm (62.6%) compared with 338 to 98 mA/mm (71.0%) for the RA channel, and a lower subthreshold swing (459 for DA versus 550 mV/dec for RA). All tested DA devices also exhibit superior threshold-voltage stability (<±10% versus >±19% for RA) and maintain an ON/OFF current ratio >$10^4$ at 800 °C, compared with ~$10^3$ for RA, thus establishing the DA channel HEMTs as a promising new platform for high-temperature electronics.**

Wide-bandgap (WBG) GaN-channel high electron mobility transistors (HEMTs) have achieved widespread commercial adoption in medium-voltage (600-1000 V) power applications [1], [2], and have also been demonstrated for high-temperature operation [3], [4], [5], [6], [7]. To meet the demand for higher operating voltages and extreme-temperature electronics, research focus has shifted toward exploring $Al_xGa_{1-x}N$ ultra-wide bandgap (UWBG, x < 0.6) and extreme bandgap (EBG, x > 0.6) channel HEMTs [8], [9], [10]. These devices exploit an exceptionally high critical electric field (>6 MV/cm) and superior transport properties of UWBG/EBG $Al_xGa_{1-x}N$ channel. Beyond high-voltage, UWBG/EBG $Al_xGa_{1-x}N$ HEMTs are also being evaluated for high-frequency [11], [12], and high-temperature applications [13], [14], [15], [16]. While silicon-based electronics are thermally constrained to operation below 300 °C and SiC devices face interface degradations above 600 °C [17], UWBG/EBG (e.g. $Al_xGa_{1-x}N$, $Ga_2O_3$, and diamond) can withstand substantially harsher thermal environments due to their large bandgaps and low intrinsic carrier concentrations [18], [19], [20], [21].

EBG $Al_xGa_{1-x}N$ channel HEMTs (x>0.6) have been reported both on AlN/sapphire templates and bulk AlN substrates [8], [9]. Nearly all the reported studies have used the $Al_xGa_{1-x}N$ random alloy channel confined between higher-Al-content $Al_yGa_{1-y}N$ barrier and back-barrier layers, typically with a substantial Al-composition offset (y-x ≥ 0.2) to provide strong carrier confinement. To facilitate low-resistance surface ohmic contact formation, all these reported devices employed thin (typically 40-80 nm) highly Si-doped, reverse composition graded $Al_xGa_{1-x}N$ contact layers (RGLs). These are selectively etched off from the access region prior to gate fabrication. The precise control needed for this etch makes this epilayer configuration difficult for large scale manufacturing processes. Over etching increases the sheet resistance and any under etching significantly increases the gate leakage current. To avoid this, recently we have reported a new epilayer configuration for $Al_{0.62}Ga_{0.38}N$ channel HEMTs [22], [23]. For our devices, the channel layer consisted of an AlN/GaN short period superlattice (SPSL) and the barrier layer was binary AlN. Regrown $n^{++}$-GaN contacts were used for source-drain ohmics to contact the 2-dimensional electron gas (2-DEG) on the digital alloy (DA) channel. In our past reported work we established peak current, gate-leakage, and reverse breakdown performance of our new DA device configuration to be equal to or better than the conventional RA channel design; however, their relative performance at extreme temperatures remains unexplored.

Several studies have demonstrated EBG $Al_xGa_{1-x}N$ channel HEMTs operating up to 500 °C; however, the ON/OFF ratio typically degrades from $\sim 10^9$ at room temperature to $\sim 10^4$ at elevated temperatures [13], [14], [15]. More recently, Lundh et al. demonstrated EBG $Al_{0.68}Ga_{0.32}N$ channel HEMT operation at 850 °C with an ON/OFF ratio exceeding $10^4$, although the peak drain current was limited to 31 mA/mm [16]. Notably, these extreme-temperature measurements were performed under $N_2$. Since the surrounding ambient (e.g., vacuum, $N_2$, air or other inert gas) can strongly influence thermally activated device degradation [24], inert environments suppress oxidation and provide limited insight into device stability under oxidizing conditions. This is particularly relevant to applications such as combustion environments, where the semiconductor surface, gate dielectric/passivation, and contact metallization are simultaneously exposed to thermal and chemical stresses [25]. Although open-air operation of EBG HEMTs has been demonstrated at ~400–600 °C [8], [10], device operation approaching 800 °C under combined thermal and oxidative stress remains largely unexplored.

In this work, we demonstrate high-temperature operation of AlN/GaN SPSL DA-channel HEMTs from room temperature to 800 °C in ambient air. The DA devices are further compared with compositionally similar $Al_{0.64}Ga_{0.36}N$ RA-channel devices fabricated and characterized under the same conditions. Over the entire temperature range, the DA devices exhibit reduced performance degradation relative to their RA counterparts, demonstrating the potential of thin SPSL digital-alloy channels as a robust platform for extreme-temperature electronics.

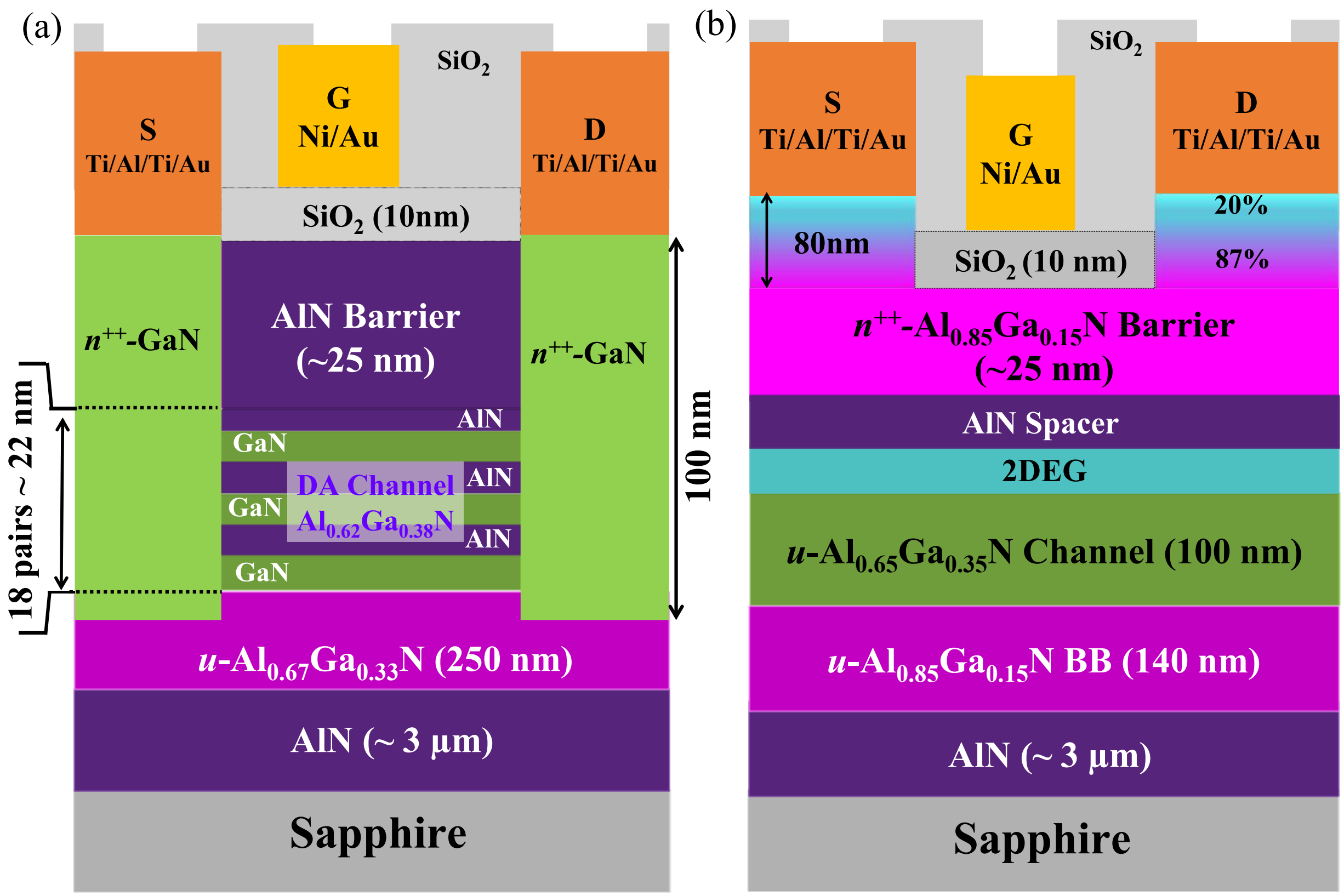

Fig. 1. (a) Schematic cross-section of the fabricated (a) AlN/GaN SPSL $Al_{0.62}Ga_{0.38}N$ digital-alloy (DA) channel (b) $Al_{0.64}Ga_{0.36}N$ random-alloy (RA) channel MOSHEMTs.

The device and epilayer structure of DA channel in this high-temperature performance study are shown in Fig. 1(a). The epilayer structure was grown using low-pressure metalorganic chemical vapor deposition (MOCVD) on AlN (thickness >2µm)/sapphire templates. It had a 250-nm undoped $Al_{0.67}Ga_{0.33}N$ back barrier, a 22-nm thin channel (18-period short-period superlattice $AlN_{7Å}$ -$GaN_{5Å}$) with an effective Al mole fraction of $x \approx 0.62$, followed by a 25-nm AlN barrier to maximize polarization-induced 2DEG density. Details of the epilayer structure are provided in our earlier report [30].

Selective-area growth (SAG) was implemented by patterning a 200-nm-thick $Si_3N_4$ hard mask and defining 70-nm-deep contact trenches via $Cl_2/BCl_3$-based inductively coupled plasma (ICP) etching, into which 100 nm of heavily Si-doped $n^{++}$-GaN was selectively regrown using pulsed MOCVD. Following mask removal, Ti/Al/Ti/Au (15/70/30/50 nm) ohmic stacks were e-beam evaporated and annealed at 950 °C for 30s in $N_2$ ambient, after which device isolation was achieved via mesa etching. A 10-nm $SiO_2$ gate dielectric was deposited by PECVD at 350 °C, followed by the definition of Ni/Au (100/200 nm) gate electrodes, Ti/Ni/Au/Ti (50/70/150/15 nm) probe pads, and a 250-nm $SiO_2$ passivation layer. The devices feature a gate length ($L_G$) of 1 µm, a source–gate spacing ($L_{SG}$) of 2 µm, and gate–drain spacings ($L_{GD}$) of 3 µm. Identical-geometry control devices were fabricated on our previously reported $Al_{0.64}Ga_{0.36}N$ random-alloy (RA) channel epilayers [3], [26], shown in Fig. 1(b), alongside the DA devices, using comparable ohmic contacts, gate metallization, gate dielectric, probe pads, and passivation. Temperature-dependent DC measurements for both types of devices were characterized under the same conditions for direct comparison using an Instec high-temperature probe station equipped with a Keysight B1500A semiconductor device analyzer. To ensure thermal equilibrium, prior to electrical characterization the devices were allowed to stabilize for 15 min at each temperature setpoint.

Device TLM I-Vs were measured from room temperature to 800 °C to extract the total contact resistance ($R_c$) and channel sheet resistance ($R_{sh}$). For the DA devices, $R_c$ includes contributions from the metal/$n^{++}$-GaN, regrown $n^{++}$-GaN and the regrown $n^{++}$-GaN/DA-channel interface. As shown in Fig. 2(a), the extracted total contact resistance exhibits minimal temperature dependence for both DA (~6 Ω·mm) and RA (~2 Ω·mm) devices over the entire temperature range. In contrast, the extracted sheet resistance increases by approximately ~5×, from 2.5 to 12.3 kΩ/□ for DA, and by ~6.5×, from 3.1 to 20 kΩ/□ for RA between 25 and 800 °C. This increase is primarily attributed to the reduction in channel mobility at the elevated temperatures. The temperature-dependent mobility for both DA and RA channel was extracted from the measured channel conductance described in our previous study of $Al_{0.62}Ga_{0.38}N$ DA-channel HEMTs [23] and the results are included in Fig. 2(b). Although mobility decreases substantially for both structures with increasing temperature, the DA channel consistently maintains higher mobility than the RA channel. At 800 °C, the DA mobility remains 29.3 $cm^2/V\cdot s$ compared with 24.2 $cm^2/V\cdot s$ for RA (~21% higher), consistent with the slower increase in $R_{sh}$ observed for the DA channel. The improved mobility of the DA channel can be attributed, at least in part, to the suppression of alloy-disorder scattering inherent to ternary random-alloy $Al_xGa_{1-x}N$ channels, while polar optical phonon (POP) scattering becomes increasingly significant at elevated temperatures [27], [28].

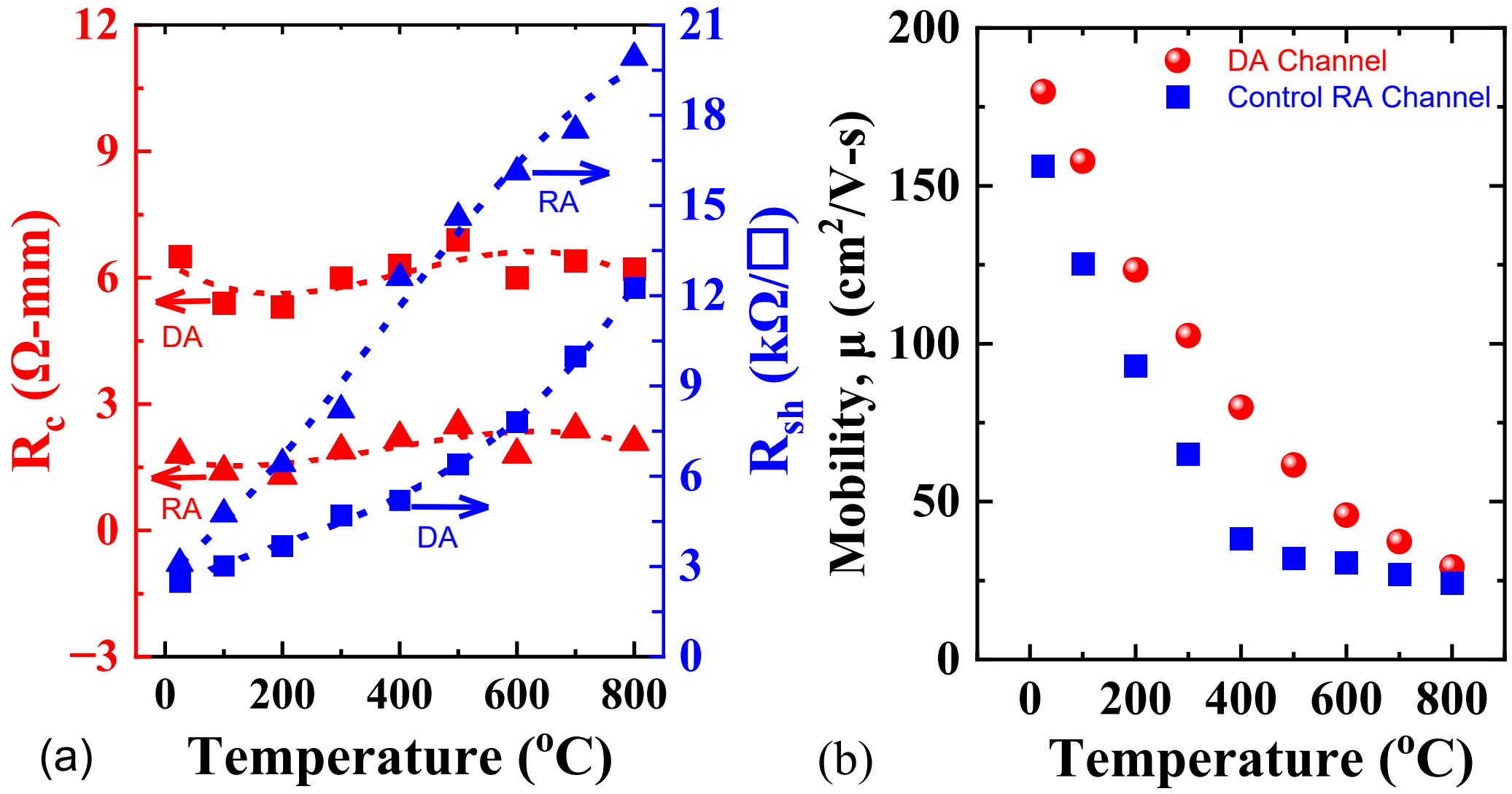

Fig. 2. (a) Extracted $R_c$ (left axis) and $R_{sh}$ (right axis), with third-order polynomial fits, and (b) Calculated Field-effect mobility for both DA and control-RA channel as a function of temperature.

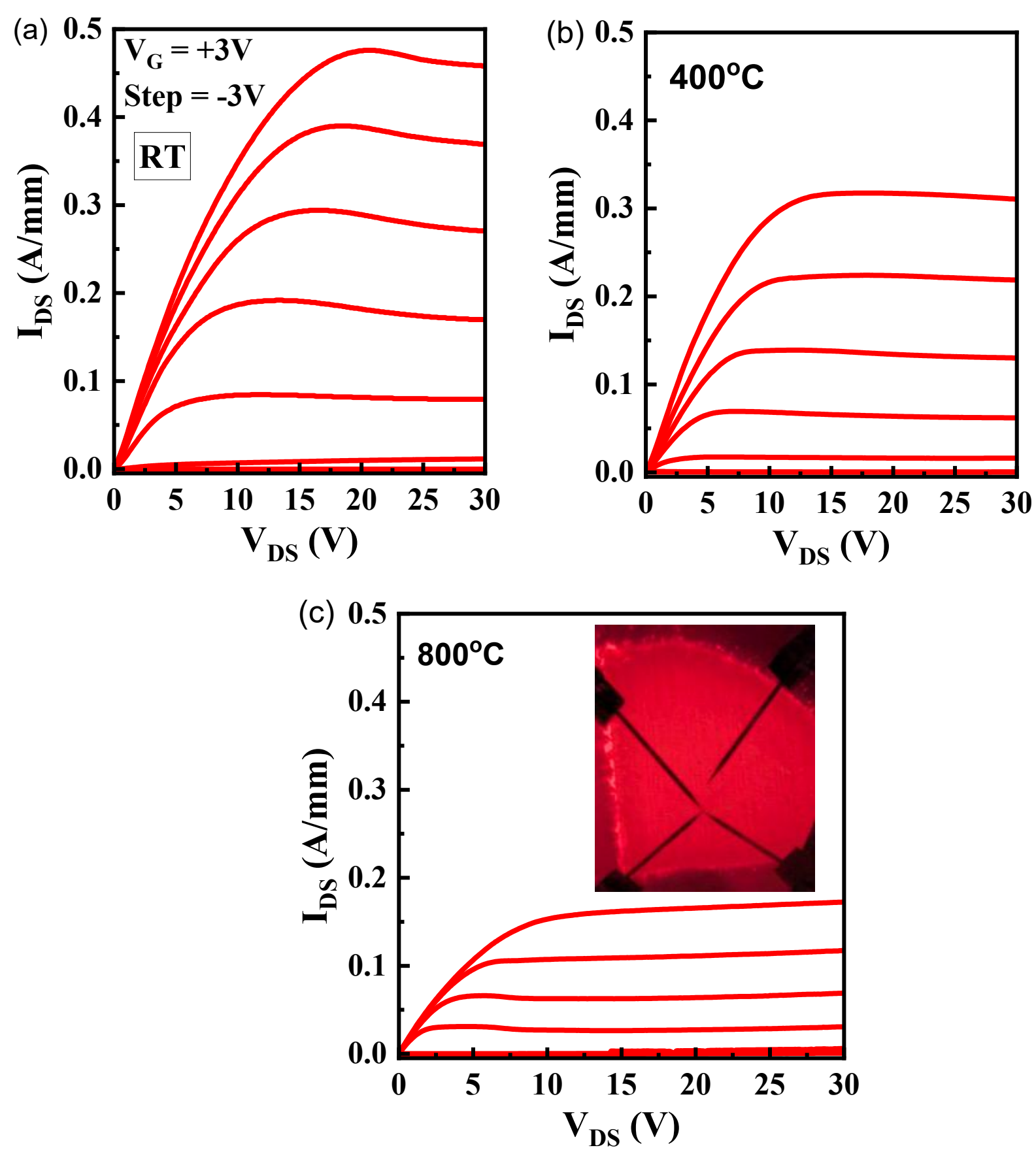

Fig. 3. Measured DC output characteristics of DA Channel MOSHEMT at (a) room temperature (RT), (b) 400 ºC, and (c) 800 ºC. The inset in (c) shows the device under measurement at 800 ºC.

Temperature-dependent DC output characteristics of DA devices are included in Fig. 3 (a)-(c). For these measurements, the drain voltage sweep was from $V_{DS}$ = 0 to 30V. For the temperature scan from room temperature (25 ºC), to 800 ºC the saturated drain current decreases from 460 mA/mm to 172 mA/mm, corresponding to a 62.6% reduction.

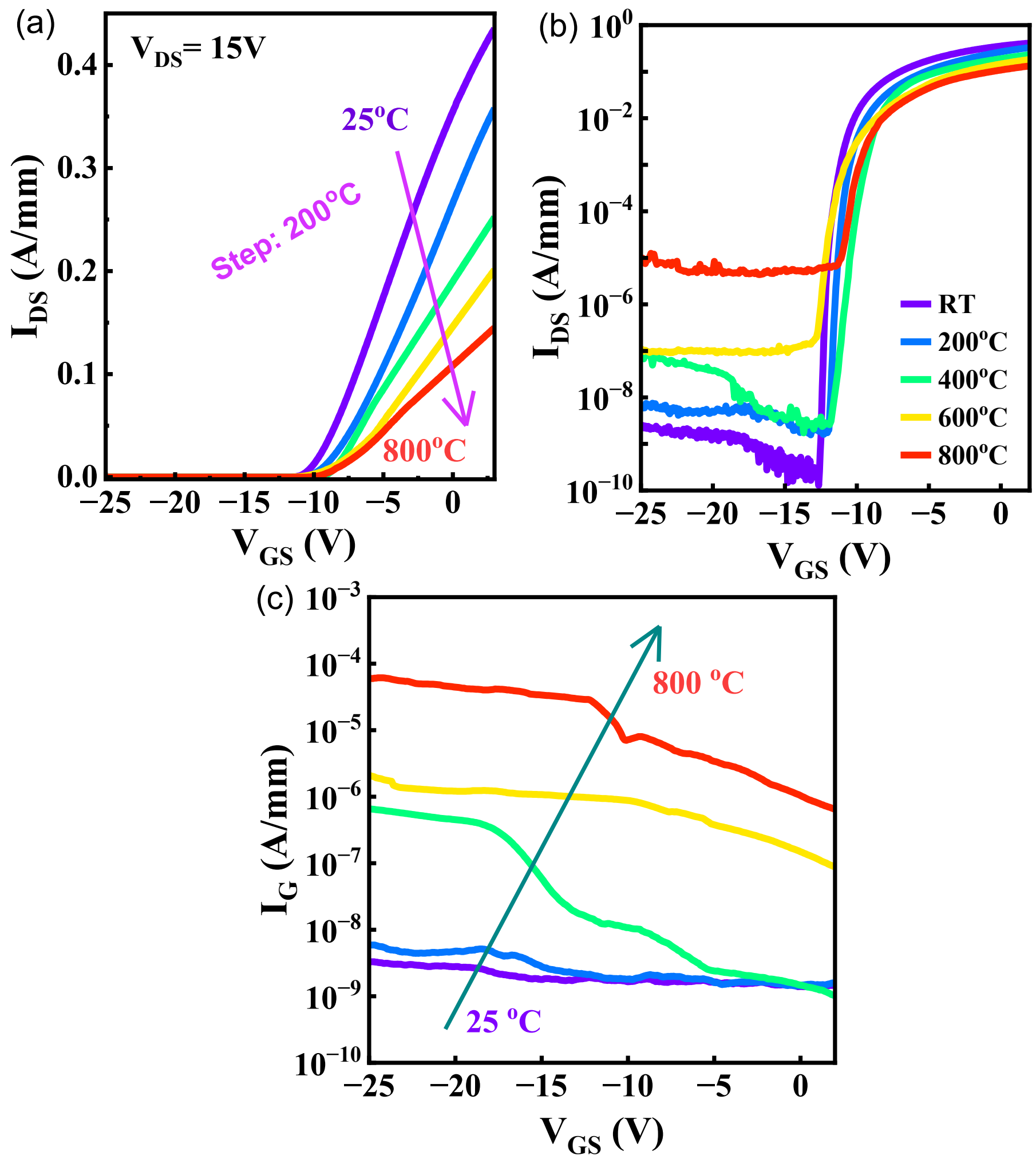

Fig. 4. Measured transfer characteristics of DA channel devices: (a) linear plot, (b) logarithmic plot, and (c) gate leakage at different temperatures.

Fig. 4(a)-(b) presents the transfer characteristics ($I_{DS}$-$V_{GS}$) of the DA-channel devices measured from 25 °C to 800 °C on linear and semi-logarithmic scales, respectively. Data are shown for operation at 25 °C, 200 °C, 400 °C, 600 °C, and 800 °C. Fig. 4(c) shows the corresponding gate leakage ($|I_G|$) characteristics measured at the same temperatures at $V_{DS}$ =15V. Despite the substantial reduction in on-state $I_{DS}$ current, and simultaneous increase in $I_G$ and off-state $I_{DS}$, the device maintains an ON/OFF current ratio exceeding $10^4$ at 800 °C.

Fig. 5 shows the variation of DA device performance parameters, namely the saturated drain current, $I_{D,sat}$ (averaged from 25 V ≤ $V_{DS}$ ≤ 30 V, $V_{GS}$ = 3 V), off-state gate current, $I_{G,off}$ (averaged from -25 V ≤ $V_{GS}$ ≤ -20 V, $V_{DS}$ = 15 V), threshold voltage, $V_{TH}$ (at $I_{DS}$ = 10 μA/mm), and ON/OFF ratio, $I_{ON}/I_{OFF}$ ($I_{ON}$ averaged from 0 V ≤ $V_{GS}$ ≤ 3 V; $I_{OFF}$ averaged from -20 V ≤ $V_{GS}$ ≤ -15 V) as a function of temperature from 25 °C to 800 °C. Data for three different devices is included to get an estimate of the variation. As shown $I_{D,sat}$ decreases with temperature up to 800 °C and this, as stated previously, is primarily due to a reduction in channel mobility. $I_{G,off}$ remains nearly constant up to about 300 °C, then increases to values included in Fig. 4 (c) as well. As seen in Fig. 5 (c), the threshold variation from 25-800 °C operation is less than ± 10%. Also, from the data in Fig. 5 (d), the device ON/OFF ratio exceeds $10^4$ even for operation at 800 °C (best ~ $6.5\times10^4$). It should also be noted that for 600 °C operation, all devices have an ON/OFF ratio exceeding $10^6$.

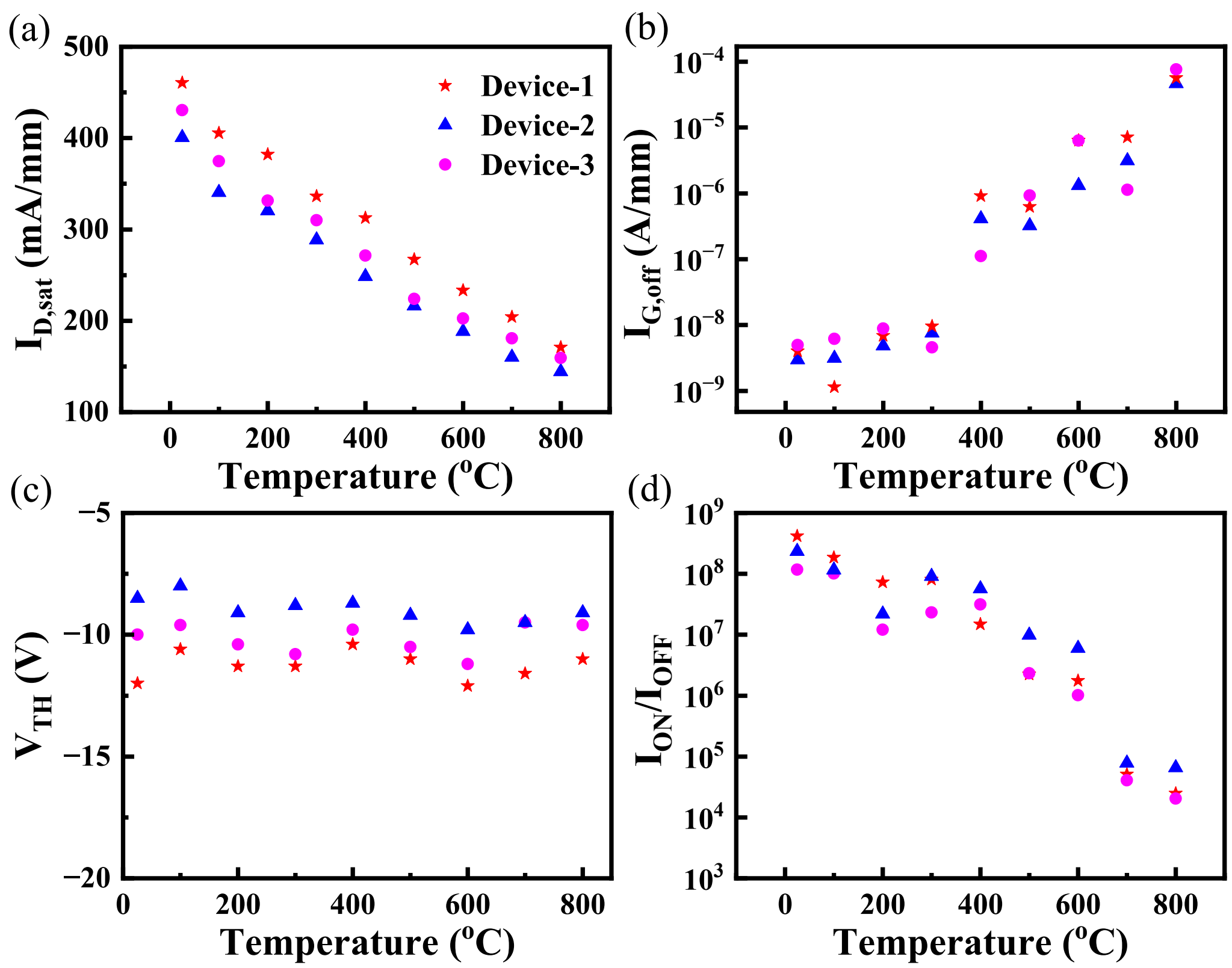

Fig. 5. Measured temperature-dependent DA Channel device metrics, including (a) $I_{D,sat}$, (b) $I_{G,off}$, (c) $V_{TH}$, and (d) ON/OFF current ratio for three representative devices.

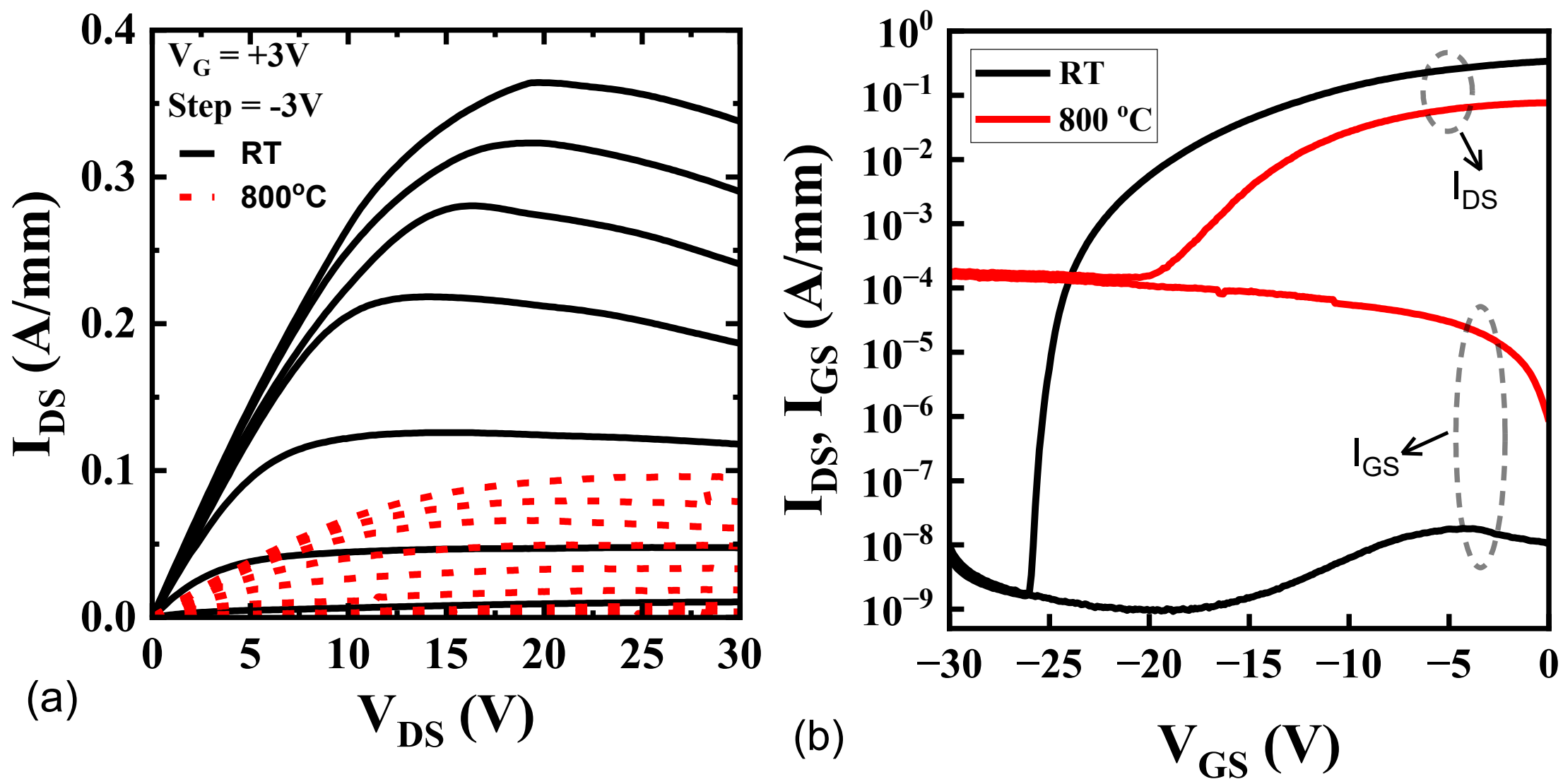

Fig. 6. Measured DC characteristics of the control $Al_{0.64}Ga_{0.36}N$ RA Channel MOSHEMT at RT and 800 °C: (a) output and (b) transfer and gate-leakage characteristics.

Fig. 6 shows the DC characteristics of the identical-geometry control RA-channel MOSHEMTs at RT and 800 °C, including (a) output characteristics and (b) transfer and gate-leakage characteristics. In comparison to the output curve of DA devices in Fig. 3, the compositionally similar RA-channel MOSHEMT exhibits a larger reduction of 71.0% (338 to 98 mA/mm). Lundh et al. reported a ~ 61% reduction for $Al_{0.68}Ga_{0.32}N$ RA-channel HEMTs under $N_2$ ambient [16]; however, their initial peak drain current at RT was only 80 mA/mm, approximately 5.8× and 4.2× lower than those of our DA and RA devices, respectively.

In Fig. 7, we have compared the temperature-dependent drain-current retention and subthreshold swing (SS) of the DA-and RA-channel MOSHEMTs. As shown in Fig. 6(a), the DA channel maintains superior performance over the entire temperature range, retaining ~37% of its room-temperature drain current at 800 °C compared with only ~30% for the RA channel. Similarly, Fig. 6(b) shows consistently lower SS for the DA channel, reaching ~459 mV/dec at 800 °C compared with ~550 mV/dec for RA, further demonstrating improved gate control at elevated temperatures. Also included for comparison are the reported values of SS for EBG $Al_{0.70}Ga_{0.30}N$

RA channel HEMTs for high temperature operation [14]. As can be seen, the DA channel device outperforms the reported RA AlGaN channel devices.

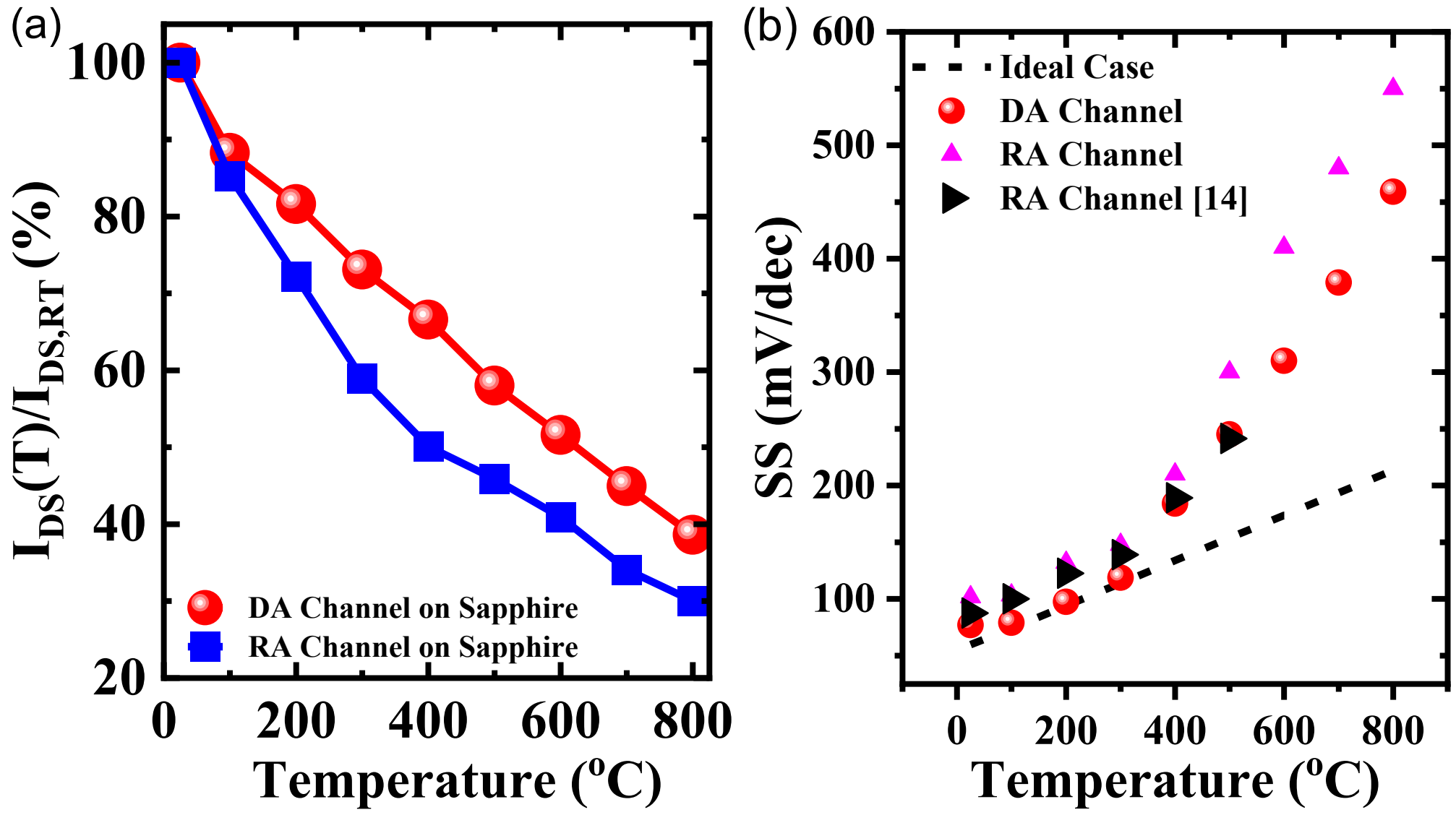

Fig. 7. High-temperature benchmarking of identical MOSHFETs with DA channel and RA channel structure: (a) $I_{DS}$ thermal stability, (b) Calculated subthreshold characteristics compared to the ideal transport limit and previously reported RA channel.

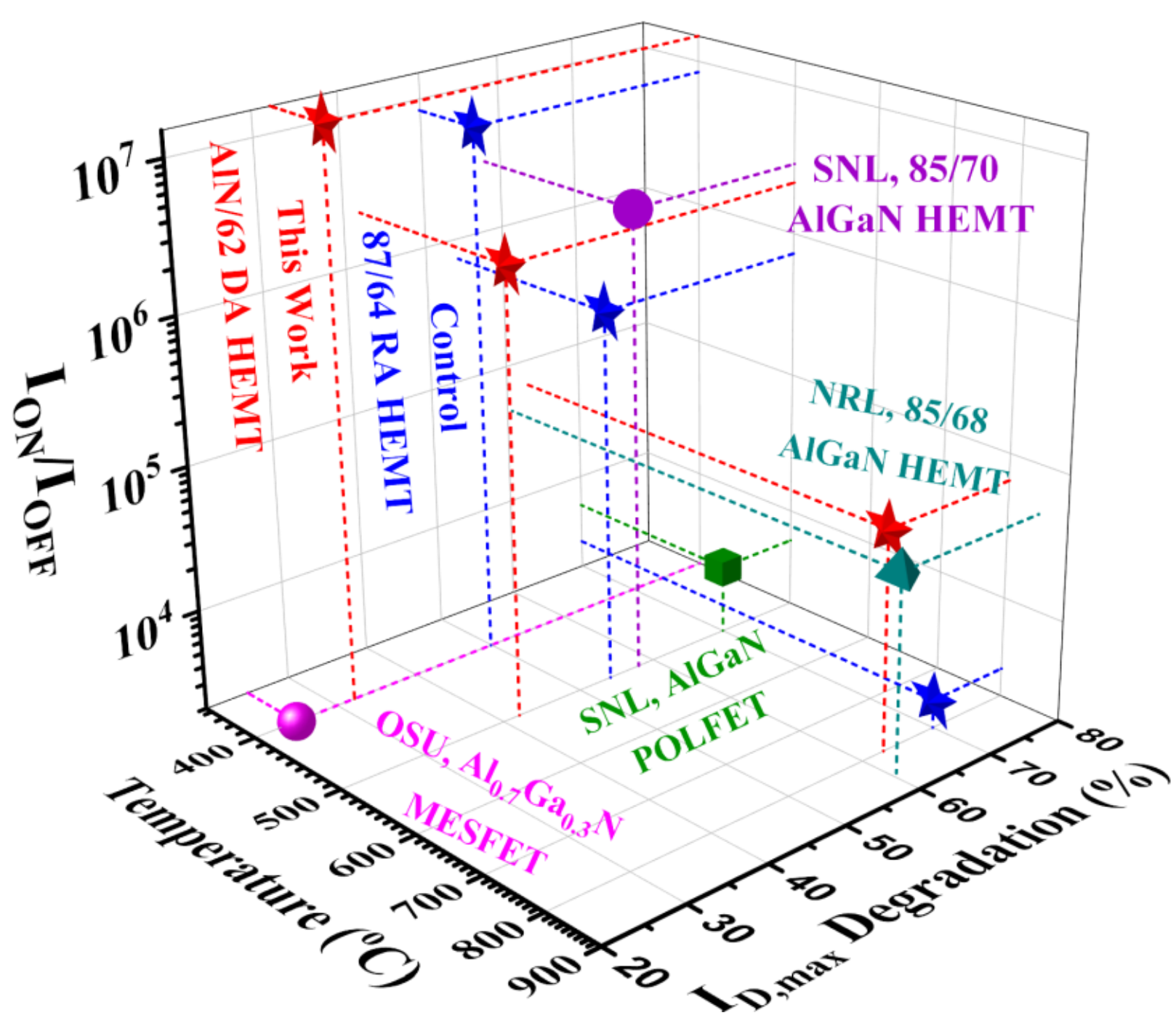

Fig. 8. Benchmarking $I_{ON}/I_{OFF}$ ratio and $I_{D,max}$ degradation (%) as a function of operating temperature for $Al_xGa_{1-x}N$ channel devices [23], [24], [25], [29].

In Fig. 8, a benchmark plot of the high-temperature performance of both our fabricated devices against previously reported EBG AlGaN-channel HEMTs is shown. The comparison includes $I_{ON}/I_{OFF}$, and the percentage of $I_{D,max}$ degradation. As seen, our DA HEMTs employing AlN/GaN SPSL channels with an average Al composition of 62% exhibit superior high-temperature performance compared with the reported devices, as well as the compositionally similar RA devices fabricated and measured alongside the DA devices, all of which employ conventional random-alloy $Al_xGa_{1-x}N$ channels.

In summary, we have demonstrated the operation of an extreme-bandgap $Al_{0.62}Ga_{0.38}N$ short-period superlattice (SPSL) digital alloy channel MOSHEMT on sapphire operating in ambient air up to 800 °C maintaining $I_{DS}$=172 mA/mm, $I_{ON}/I_{OFF}$ >$10^4$, and VTH variation within ±10%. Compared with compositionally similar $Al_{0.64}Ga_{0.36}N$ RA-channel devices, the DA architecture exhibited higher drain-current retention (37.4% versus 29.0%), higher mobility (29.3 versus 24.2 $cm^2/V\cdot s$), lower SS (459 versus 550 mV/dec) at 800 °C. These results establish AlN/GaN SPSL digital-alloy channel engineering as a robust, oxidation-tolerant platform for extreme-temperature III-nitride electronics; closing a critical gap between the inert-ambient demonstrations reported to date and the combined thermal and chemical stresses of real-world high-temperature environments.

**Acknowledgment**

This work was supported by ADI under award number 10015113, ONR grant number N00014-23-1-2010 (Capt. Lynn Petersen) and N00014-23-1-2289. It was partially supported by the AFOSR award number FA9550-24-1-0269, and by the USC Research Institute for Extreme Electronics grant number 80005646. Author(s)' opinions, findings, conclusions, and recommendations may not necessarily represent those of the Department of War or the U.S. Government. This work was also supported under activities funded by the Microelectronics Commons Program, a DoW initiative, under award number N00164-23-9-G060.

## AUTHOR DECLARATIONS

### Conflict of Interest

The authors have no conflicts to disclose.

**DATA AVAILABILITY**

The data that support the findings of this study are available from the corresponding author upon reasonable request.